\documentclass[aps,twocolumn,amssymb,showpacs,footinbib]{revtex4}

\newcommand{\be}{\begin{equation}}
\newcommand{\ee}{\end{equation}}
\newcommand{\bea}{\begin{eqnarray}}
\newcommand{\eea}{\end{eqnarray}}

\begin{document}

\title{Stress tensor for extreme 2D dilatonic Reissner-Nordstr\"om black holes}
\author{Serena Fagnocchi }
\altaffiliation{Email addresses: fagnocchi@bo.infn.it}
\affiliation{Dipartimento di Fisica dell'Universit\`a di Bologna
and INFN sezione di Bologna, \\  Via Irnerio 46,
40126 Bologna, Italy }
\author{Sara Farese}
\altaffiliation{Email addresses: farese@ific.uv.es}
\affiliation{Departamento de F\'{\i}sica Te\'orica and IFIC, Centro Mixto Universidad de Valencia-CSIC,\\
Facultad de F\'{\i}sica, Universidad de Valencia,\\
Burjassot-46100, Valencia, Spain.}

\begin{abstract}
We calculate the expectation value of the stress energy tensor for a
massless dilaton-coupled 2D scalar field propagating on an
extremal Reissner-Nordstr\"om black hole formed by the collapse of
a timelike shell, showing its regularity on the horizon.
\end{abstract}


\maketitle

Extremal black holes are peculiar objects which enter  various and
important contexts of gravitational physics. Their interest is
related to the fact that, being characterized by a zero Hawking
temperature, they represent the natural end-state configuration of
the evaporation of non extremal charged black holes.  It would be
rather disappointing if these configurations turned out to be
singular.
 In
a previous work \cite{BFFFN-S} we calculated, within the Polyakov
theory,  the expectation value of the energy-momentum tensor for a
2D massless  scalar field propagating on an extreme
Reissner-Nordstr\"om black hole, showing, unlike previous results
\cite{Trivedi,Loranz...}, regular behaviors on the horizon. The
problem was treated considering the process of formation of an
extremal black hole through the collapse of a charged null shell.
Our purpose here is to generalize the previous work to show that
the basic result does not depend on the specific model chosen. To
this end we shall consider the more general case of a dilaton
coupled scalar field theory, and  the formation of the hole
through the collapse of a timelike shell. This is  a non trivial
extension with respect to the pure 2D Polyakov theory, since new
divergent terms due to the presence of the dilaton
 appear; they look quite different from the
ones emerging in Polyakov theory and their regularity is not obvious.\\
In quantum field theory in curved spacetimes  the mean value of
the matter fields energy momentum tensor plays a fundamental role
since it determines the behavior of the solutions of the
semiclassical evolution equations. The main problem is then to
find a covariant expression for this quantum stress tensor. This
is a very difficult task for an arbitrary 4D spacetime, where no
analytical expression for $T_{\mu\nu}$ exists. Nevertheless,
starting from a 4D spherically symmetric theory, one can perform a
dimensional reduction leading to a 2D model, where detailed
predictions can be made. In the two dimensional theory obtained in
this way, the field is coupled not only to the 2D metric, but also
to the dilaton  $\phi$,  which is linked to the radius of the
two-sphere and reminds the
four-dimensional origin of the theory we started from. \\
Let us consider a four-dimensional massless minimally coupled scalar
field with action
\be \label{4daction}
S^{(4)}=-\frac{1}{8\pi}\int
d^4x\sqrt{-g^{(4)}}(\nabla f)^2.
\ee
A spherically symmetric  four-dimensional
metric can be written as
\be \label{sphsymetric}
ds_{(4)}^2=ds_{(2)}^2+r_0^2 e^{-2\phi}d\Omega^2,
\ee
where we have
parameterized the radius of the two-sphere $r^2=r_0^2 e^{-2\phi}$
through a dilaton field $\phi$. $r_0$ is a scale factor which can
be set  to 1 without loss of generality. We assume that
the field $f$ is a function of the $(t,r)$ coordinates only. Since
any field in a spherically symmetric background can be expanded in
spherical harmonics, this hypothesis corresponds to picking up
only the s-wave component. Through the hypothesis of spherical
symmetry for both the metric $g_{\alpha\beta}$ and the field $f$,
the 4D action (\ref{4daction}) can be integrated with respect to the
angular coordinates, to obtain the two-dimensional action
\be
\label{2daction}
S^{(2)}=-\frac{1}{2}\int
d^2x\sqrt{-g^{(2)}}e^{-2\phi}(\nabla f)^2,
\ee
 where $g^{(2)}$
represents the metric in the $(t,r)$ plane. This shows that in the 2D theory the scalar field acquires a non trivial
coupling with the dilaton field besides the usual one to the metric. \\
Since every  two-dimensional spacetime is conformally flat, it is
always possible to write the line element in conformal coordinates
$\{{x^\pm}\}$ as: \be ds_{(2)}^2=-e^{2\rho(x)}dx^+ dx^-. \ee It is
worth noting that this choice is not unique, but infinite sets of
conformal coordinates can be obtained from $\{{x^\pm}\}$ through
conformal
transformations.\\
Let us briefly remind  the expression for the expectation value of the
covariant quantum stress tensor for this 2D dilatonic theory in a
generic quantum state $|\Psi\rangle$ \cite{FFN-S}. Using the anomalous
transformation law of the normal ordered stress tensor combined with
  equivalence principle argument (for details see \cite{FFN-S,FN-S}), one gets
\bea \label{qstdilaton3}
\langle\Psi|T_{\pm\pm}(x^+,x^-)|\Psi\rangle
&=& \langle\Psi|:T_{\pm\pm}(x^+,x^-):|\Psi\rangle+\nonumber\\
&&-\frac{1}{12\pi}(\partial_{\pm}\rho\partial_{\pm}\rho -
\partial_{\pm}^2\rho)  +\\ &&+ \frac{1}{2\pi} \left[  \partial_{\pm}\rho
\partial_{\pm}\phi + \rho (\partial_{\pm}\phi)^2 \right]. \nonumber
\eea
Note that the covariant stress tensor (l.h.s of the previous
relation) is given by its corresponding normal ordered tensor
calculated in the same quantum state $|\Psi\rangle$, plus some
local terms, the last of which depends on the dilaton field. The
difference with respect to the minimally coupled case is
represented only by the last terms induced by the dilaton, since
the others give exactly  the Polyakov stress tensor \cite{Polyakov}.\\
It can also be shown that, imposing energy conservation, a
state-independent $\langle T_{+-} \rangle$ component appears \cite{FN-S}:
\be\label{trace}
\langle T_{+-}
\rangle=-\frac{1}{12\pi}(\partial_+\partial_-\rho+3\partial_+\phi\partial_-\phi-3\partial_+\partial_-\phi).
\ee
This is a pure quantum effect, since the classical theory is
Weyl invariant, and eq.(\ref{trace}) is in agreement with the well known results about trace
anomaly \cite{Muk}.\\
In analogy with the Polyakov case, from eq.(\ref{qstdilaton3})  one can find that the relation
between the expectation value of the stress tensor in two
different vacuum states  as
 \bea \label{transf2}
\langle \tilde{x}^\pm|T_{\pm\pm}(x^+,x^-)|\tilde{x}^\pm\rangle&=&
\langle
x^\pm|T_{\pm\pm}(x^+,x^-)|x^\pm\rangle+\nonumber\\
&&-\frac{1}{24\pi}
\{ \tilde{x}^{\pm}, x^{\pm} \}+\\
&&+\frac{1}{4\pi}
\ln\bigg(\frac{dx^-}{d\tilde{x}^-}\frac{dx^+}{d\tilde{x}^+}\bigg)(\partial_\pm \phi )^2+\nonumber\\
&&+\frac{1}{4\pi}\frac{d^2x^\pm}{d\tilde{x}^{\pm2}}\bigg(\frac{dx^\pm}{d\tilde{x}^\pm}\bigg)^{-2}
\partial_{\pm}\phi,\nonumber\eea
where $|x^\pm\rangle$ represents the vacuum state determined by
the expansion of the field in the modes which  are positive
frequency  with respect to the time $(x^++x^-)/2$. Analogously
$|\tilde{x}^\pm\rangle$ is the vacuum associated to the modes
positive frequency with respect to the time
$(\tilde{x}^++\tilde{x}^-)/2$. Finally \be \{ \tilde{x}^{\pm},
x^{\pm}
\}=\frac{d^3\tilde{x}^\pm}{dx^{\pm3}}\big/\frac{d\tilde{x}^\pm}{dx^\pm}
-\frac{3}{2}\bigg(\frac{d^2\tilde{x}^\pm}{dx^{\pm2}}/\frac{d\tilde{x}^\pm}{dx^\pm}\bigg)^2
\ee is the Schwarzian derivative associated to the conformal
transformation ${x^\pm}\to{\tilde{x}}^\pm$. In eq.(\ref{transf2})
the difference with the Polyakov case is represented by the last
local terms induced by the dilaton field.\footnote{It is possible
to face the problem from a different point of view, choosing as
effective action for the quantum fields the one obtained by
functional integration of the trace anomaly (the so  called
"anomaly induced effective action" $S_{aind}$ \cite{Muk,BF}).
However this approximation for the effective action is in general
not reliable since it predicts an unphysical negative energy
outflux for the Schwarzschild black hole \cite{Muk,BF}. In any
case also in these models one can construct a regular $T_{\mu\nu}$
on the horizon of extremal RN background.}
\\
Clearly, the $\langle T_{+-}\rangle$ components in the
two vacuum states coincide:
\be
 \langle \tilde{x}^\pm| T_{+-} (x^+,x^-)
|\tilde{x}^\pm\rangle=\langle x^\pm|T_{+-}(x^+,x^-)|x^\pm\rangle.
\ee
Now we want  to apply this formalism to the 2D extremal
Reissner-Nordstr\"om spacetime, whose line element is given by \be
\label{metric} ds^2=-f(r)dt^2 + \frac{dr^2}{f(r)} =-f(r)dudv \,
\ee where $f(r)=(1-M/r)^2$. Extreme black holes are charged
Reissner-Nordstr\"om black holes whose charge assumes its maximum
value, i.e. $|Q|=M$ and possesses a single degenerate horizon at
$r=M$ with zero surface gravity. $u$ and $v$ are respectively the
retarded and advanced Eddington-Finkelstein coordinates \be
u=t-r^*\ , \ \ \ v=t+r^* \ee where \be r^\ast=\int
\frac{dr}{\big(1-\frac{M}{r}\big)^2}=r+2M\ln\bigg(\frac{r}{M}-1\bigg)-\frac{M^2}{(r-M)}.
\ee In the literature it has been shown that a static $\langle
T_{\mu\nu}\rangle$ calculated in this background is divergent on
the horizon \cite{Trivedi}. In a recent paper \cite{BFFFN-S} it
was shown how such diverging behavior completely disappears in the case of a collapse if
one takes into account the time-dependent contribution to  $\langle
T_{\mu\nu}\rangle$ induced by the collapse itself.
 Incoming modes,
which are asymptotically $1/r e^{-i\omega v}$, reflected through the origin, emerge  just before the horizon
formation transformed into complicated redshifted modes
which are positive frequency with respect to a Kruskal retarded coordinate $U$ \cite{BirrellDavies}. The relation between the $u$ and
the Kruskal $U$ coordinate is given at late times as
\be
 u=-4M \bigg[\ln\bigg(-\frac{U}{M}\bigg)+\frac{M}{2U}\bigg].
\label{u}
\ee
This was obtained in the case of a colapsing null shell \cite{BFFFN-S}.\\
Now we are going to show how this result is completely
independent on the details of the collapse. To  this aim we will
perform the same
calculation for a timelike shell and we will show that indeed we recover the same result.\\
Outside the collapsing shell the line element is
\bea ds^2&=&-f(r)dudv,\\
u&=&t-r^\ast+R_o^\ast,\\
v&=&t+r^\ast-R_o^\ast,\\
R_o^\ast&=&const,
\eea
and the dilaton is linked to the radial coordinates through
\be
\phi=-\ln r\ .
\ee
Inside the shell instead
\bea ds^2&=&- dUdV, \\
\label{U2}
U&=&\tau-r+R_0,\\
\label{V2}
V&=&\tau+r-R_0,
\eea
where the relation between $R_0$ and $R_o^\ast$ is the same as that
between $r$ and $r^\ast$.\\
Before $\tau=0$ the shell is at rest with its surface at $r=R_0$,
while for $\tau>0$ the shell will follow the world line $R(\tau)$.
Matching the inner and the outer metrics along the collapsing
shell (for details see \cite{BirrellDavies}), in the near horizon
limit we finally find \be \label{alphaprimo2} \frac{dU}{du}\sim
\frac{(\dot{R}-1)}{2\dot{R}}f(R) \ee where the dot represents
differentiation with respect to $\tau$.
\\
Let us expand $R(\tau)$ around the value it assumes on the
horizon: \be\label{R_tau2}
R(\tau)=R_H+\dot{R}_H(\tau-\tau_H)+\frac{1}{2}\ddot{R}_H(\tau-\tau_H)^2+O((\tau-\tau_H)^3),
\ee where the suffix $H$ means the value the function assumes on
the horizon. Note that $\dot{R}_H<0$ since the ball
is shrinking.\\
We need to find the explicit form for the transformations between the
$\{u,v\}$ and $\{U,V\}$ coordinates in the near horizon limit. To
solve eq.(\ref{alphaprimo2}) it is necessary also to expand $f(R)$
for $R\rightarrow R_H$: observe that a double pole for $R=R_H=M$
appears. From $U=\tau-R(\tau)+R_0$ and
 eq.(\ref{R_tau2}) we get
\be
f(U)\simeq\frac{\dot{R}^2_HU^2}{M^2(1-\dot{R}_H)^2+2M\dot{R}_H(1-\dot{R}_H)
U+aU^2}. \ee with $a=(\dot{R}_H^2+M\ddot{R}_H)$. If we insert the
above approximation into (\ref{alphaprimo2}), a straightforward
integration gets \be \label{cu} u=-4M
\bigg[\ln\bigg(-\frac{U}{cM}\bigg)+\frac{cM}{2U}\bigg]
\qquad\mbox{for}\,U\to 0 \ee with $c=
\frac{\dot{R}-1}{\dot{R}}\big|_{R_H}$ and an opportune choice of
the integration constant. Note that eq.(\ref{cu}) has the same
form of the coordinate transformation given in eq.(\ref{u}). More
precisely the two equations  exactly coincide if we just rescale
$U/c\to U$. This difference however will not affect the
calculation we are facing, so, without loss of generality, in the
following we will
use eq.(\ref{u}).\\
Now using eq.(\ref{transf2}) we can  compute the stress energy tensor expectation value at late time on the
$|U\rangle$  vacuum.   First we shall find the expectation value of the
stress tensor on the Boulware state
$|B\rangle$. Eq.(\ref{qstdilaton3}), with
$\{x^\pm\}=\{u,v\}$ the Eddington-Finkelstein coordinates, yields
\bea \ \langle B|T_{uu}|B\rangle &=&\langle
B|T_{vv}|B\rangle = \nonumber\\
&=&-\frac{1}{24\pi}\frac{M}{r^3}\bigg(1-\frac{M}{r}\bigg)^3 +\label{Boulware1}\\
&&-\frac{1}{8\pi}\frac{M}{r^3}\bigg(1-\frac{M}{r}\bigg)^3 +\frac{1}{16\pi
r^2}f^2\ln f=\nonumber\\
&=&-\frac{1}{6\pi}\frac{M}{r^3}\bigg(1-\frac{M}{r}\bigg)^3
+\frac{1}{16\pi r^2}f^2\ln f. \eea The first term in the r.h.s of
eq.(\ref{Boulware1}) represents just the expectation value of the
stress tensor in the Boulware state for the minimally coupled case
(cfr.\cite{BFFFN-S}), while the other two terms are those induced by the dilaton field.\\
To obtain the Unruh state expectation values  we can apply eq.(\ref{transf2})  with
$|\tilde{x}^\pm\rangle=|U\rangle\equiv |U,v\rangle$ and $|x^\pm\rangle=|B\rangle\equiv |u,v\rangle$.
We find:
 \bea
\langle
U |T_{uu}|U\rangle&=&
\langle B|T_{uu}|B\rangle-\frac{1}{24\pi}
\{ U,u \}+\nonumber\\
&&+\frac{1}{4\pi}
\ln\bigg(\frac{du}{dU}\bigg)(\partial_u \phi )^2+\\
&&+\frac{1}{4\pi}\frac{d^2u}{dU^2}\bigg(\frac{du}{dU}\bigg)^{-2}
\partial_{u}\phi.\nonumber\\
\langle U |T_{vv}|U\rangle &=& \langle B|T_{vv}|B\rangle  \eea
with $du/dU$ calculated from eq.(\ref{u}).\\ To
check the regularity of the stress tensor on the future horizon
$H^+$, it is necessary to express it in a  frame regular
there. The relevant component we need to compute is \bea
\label{TUU} \langle U|T_{UU}|U\rangle
&=&\bigg(\frac{du}{dU}\bigg)^2
\langle U|T_{uu}|U\rangle=\nonumber\\
&=&\frac{16M^2}{U^2}\bigg(1-\frac{M}{2U}\bigg)^2\cdot\nonumber\\
&&\cdot \bigg\{-\frac{1}{24\pi}\frac{M}{r^3}\bigg(1-\frac{M}{r}\bigg)^3+\nonumber\\
&&-\frac{1}{8\pi}\frac{M}{r^3}
\bigg(1-\frac{M}{r}\bigg)^3+\\
&&+\frac{1}{16\pi r^2}f^2\ln f +\nonumber\\
&&+\frac{1}{24\pi}\frac{U^3(U-2M)}{2M^2(2U-M)^4}+\nonumber\\
&&+\frac{1}{16\pi
r^2}f^2\ln\bigg[-\frac{4M}{U}\bigg(1-\frac{M}{2U}\bigg)\bigg]+\nonumber\\
&&+\frac{1}{4\pi}\frac{U(U-M)}{M(2U-M)^2}\frac{f}{2r}\bigg\}.\nonumber
\eea To better appreciate the difference between this and the non
dilatonic theory, we write down also the $\langle T_{UU}^P\rangle$
obtained in the context of the Polyakov theory: \bea
 \langle U|T_{UU}^P|U\rangle
&=&\bigg(\frac{du}{dU}\bigg)^2
\langle U|T_{uu}^P|U\rangle=\nonumber\\
&=&\frac{16M^2}{U^2}\bigg(1-\frac{M}{2U}\bigg)^2\cdot \nonumber\\
&&\cdot \bigg\{-\frac{1}{24\pi}\frac{M}{r^3}\bigg(1-\frac{M}{r}\bigg)^3\\
&&+\frac{1}{24\pi}\frac{U^3(U-2M)}{2M^2(2U-M)^4}\bigg\}.\nonumber
\eea So eq.(\ref{TUU}) can be rewritten as \bea \label{DvsP}
\langle
U|T_{UU}|U\rangle &=&\langle U|T_{UU}^P|U\rangle+\nonumber\\
&&\frac{16M^2}{U^2}\bigg(1-\frac{M}{2U}\bigg)^2
\cdot\nonumber\\
&&\cdot \bigg\{-\frac{1}{8\pi}\frac{M}{r^3}
\bigg(1-\frac{M}{r}\bigg)^3+\nonumber\\
&&+\frac{1}{16\pi r^2}f^2\ln f +\\
&&+\frac{1}{16\pi
r^2}f^2\ln\bigg[-\frac{4M}{U}\bigg(1-\frac{M}{2U}\bigg)\bigg]+\nonumber\\
&&+\frac{1}{4\pi}\frac{U(U-M)}{M(2U-M)^2}\frac{f}{2r}\bigg\}.\nonumber
\eea In \cite{BFFFN-S} we have already shown how the behavior of
$\langle T_{UU}^P\rangle$ is regular on the future horizon.
Eq.(\ref{DvsP}) shows clearly that new divergent terms (both
polynomial and logarithmic) in the near horizon limit $U\sim
-(r-M)$ appear. We can see  how these divergences compensate
exactly each other to give a finite net result. For the polynomial
terms we have \bea
&&\frac{16M^2}{U^2}\bigg(1-\frac{M}{2U}\bigg)^2\bigg\{-\frac{1}{8\pi}\frac{M}{r^3}
\bigg(1-\frac{M}{r}\bigg)^3+\nonumber\\
&&+\frac{1}{4\pi}\frac{U(U-M)}{M(2U-M)^2}\frac{f}{2r}\bigg\}\sim \nonumber\\
&& \sim\frac{4M^2(M-2r)^2}{(r-M)^4}\bigg\{-\frac{1}{8\pi}\frac{M(r-M)^3}{r^6}+\nonumber\\
&&+\frac{1}{8\pi}\frac{(r-M)^3}{Mr^2(M-2r)^2}\bigg\}=\nonumber\\
&&=\frac{1}{2\pi}\frac{M(r-M)(r^2+2Mr-M^2)}{r^6}\stackrel{r=M}{\longrightarrow}0.
\eea Analogously the logarithmic terms sum up to give a
non diverging result: \bea
&&\frac{16M^2}{U^2}\bigg(1-\frac{M}{2U}\bigg)^2\bigg\{\frac{1}{16\pi r^2}f^2\ln f +\nonumber\\
&&+\frac{1}{16\pi
r^2}f^2\ln\bigg[-\frac{4M}{U}\bigg(1-\frac{M}{2U}\bigg)\bigg]\bigg\}\sim\nonumber\\
&&\sim \frac{1}{4\pi}\frac{M^2(M-2r)^2}{r^6} \big\{ 2\ln
(1-\frac{M}{r})+\nonumber\\
&&+\ln \left[\frac{4M}{r-M}(1+\frac{q}{2(r-M)}) \right]
\big\}=\nonumber\\
&&=\frac{1}{4\pi}\frac{M^2(M-2r)^2}{r^6}\ln
\frac{2M(2r-M)}{r^2}\nonumber\\
&&\stackrel{r=M}{\longrightarrow}\frac{1}{4\pi M^2}\ln 2< \infty.
\eea
Near the future horizon the limit finally is
 \be
\langle U|T_{UU}|U\rangle
\stackrel{r\to M}{\longrightarrow}\frac{1}{4\pi
M^2}\bigg(\ln2-1\bigg)<\infty
\ee
that is indeed regular.\\
Had we used eq.(\ref{cu}) instead of eq.(\ref{u}), we would have
obtained a different final value for the limit:
\be
 \langle
U|T_{UU}|U\rangle \stackrel{r\to M}{\longrightarrow}\frac{1}{4\pi
c^2M^2}\bigg(\ln2-1\bigg)+O(\ln c) \ee
 which is finite too. Note
that the finite part  depends, through the constant $c$,  on the
details of the collapse without affecting the validity of our
arguments.\footnote{It is also  possible to find a new coordinate
$V$ regular on the past horizon $H^-$, $v=4M
\big[\ln\big(\frac{V}{M}\big)-\frac{M}{2V}\big]$. Using this $V$
as a Kruskal coordinate one can show that the expectation value of
$T_{VV}$
 on $H^-$ is regular. This result holds for any transformation $v\rightarrow V$
 which reduces to the previous one in the limit $V\rightarrow 0$.}\\
So the potentially diverging terms induced by the dilaton field in
the stress energy tensor compensate each other, leading to a
regular result. No physically unacceptable behavior arises.

\section*{Acknowledgments} We wish to thank R.Balbinot for essential
collaboration. We also thank J.Navarro-Salas and A. Fabbri  for
useful discussions and indications.

\end{document}